\documentclass[fleqn,10pt]{SelfArx} 

\usepackage[english]{babel} 

\usepackage{lipsum} 

 \definecolor{color1}{RGB}{0,0,110} 
\definecolor{color2}{RGB}{0,20,80} 

\usepackage{hyperref} 
\hypersetup{hidelinks,colorlinks,breaklinks=true,urlcolor=color2,citecolor=color1,linkcolor=color1,bookmarksopen=false,pdftitle={Title},pdfauthor={Author}}

\JournalInfo{Uploaded to arXiv.org} 
\Archive{Results shared on figshare.com} 

\PaperTitle{A systematic Monte Carlo simulation study of the primitive model electrical double layer over an extended range of concentrations, electrode charges, cation diameters and valences} 

\Authors{M\'onika Valisk\'o\textsuperscript{1}, Tam\'as Krist\'of\textsuperscript{1},  Dirk Gillespie\textsuperscript{2},  Dezs\H{o} Boda\textsuperscript{1,3}*} 
\affiliation{\textsuperscript{1}\textit{Department of Physical Chemistry, University of Pannonia,  P. O. Box 158, H-8201 Veszpr\'em, Hungary}} 
\affiliation{\textsuperscript{2}\textit{Department of Physiology and Biophysics, Rush University Medical Center, Chicago, IL 60612}} 
\affiliation{\textsuperscript{3}\textit{Institute of Advanced Studies K\"oszeg (iASK), Chernel u. 14, H-9730 K\"oszeg, Hungary}} 
\affiliation{*\textbf{Corresponding author}: dezsoboda@gmail.com} 

\newcommand{\keywordname}{Keywords} 

\Abstract{The purpose of this study is to provide data for the primitive model of the electrical double layer, where ions are modeled as charged hard spheres, the solvent as an implicit dielectric background (with dielectric constant $\epsilon= 78.5$), and the electrode as a smooth, uniformly charged, hard wall.
We use canonical and Grand Canonical Monte Carlo simulations to compute the concentration profiles, from which the electric field and electrostatic potential profiles are obtained by solving Poisson's equation.
We report data for an extended range of parameters including 1:1, 2:1, and 3:1 electrolytes at concentrations $c=0.0001-1$ M near electrodes carrying surface charges up to  $\sigma=\pm 0.5$ Cm$^{-2}$.
The anions are monovalent with a fixed diameter $d_{-}=3$ {\AA}, while the charge and diameter of cations are varied in the range $z_{+}=1$, 2, 3 and $d_{+}=1.5$, 3, 6, and 9 {\AA} (the temperature is $298.15$ K).
We provide all the raw data in the Supporting Information.}

\begin{document}

\flushbottom 

\maketitle 

\tableofcontents 

\thispagestyle{empty} 


\section*{Introduction} 
\label{sec:intro}
\addcontentsline{toc}{section}{Introduction} 

Electrical double layers (DLs) formed by ions near a charged surface are everywhere, from everyday technologies (e.g., near electrodes in batteries) to biological cells (e.g., near cell membranes and around proteins and DNA) to lab devices and materials (e.g., inside porous Nafion and nanofluidic devices), to name just a few.  
And while DLs have been studied for more than a century starting with the classical theory of Gouy-Chapman-Stern \cite{gouy-jp-1910,chapman_pm_1913,stern_ze_1924}, DLs have become more important than ever because of new technological applications that exploit the physics and energetics of DLs.  
For example, the capacitive energy stored in DLs forms the basis of supercapacitors (a.k.a.\ electrochemical capacitors) \cite{simon_nature_2008} and similarly capacitively holding ions near a charged surface is used for desalination \cite{porada_pms_2013}.
Temporary protection of inner metal surfaces of industrial equipments from corrosion (e.g., in oil refinery operations) is highly affected by the properties of DLs.   
In these applications, the ideal ions are those that provide a high capacitance (i.e., a lot of charge stored with little applied voltage).  
Other potential applications like optimizing sensors that read the binding and unbinding of charged aqueous ligands \cite{khan_am_2010} utilize ions that provide low capacitance \cite{friedrich_sab_2016}.  
DLs can also be used to efficiently convert pressure into voltage in nanofluidic devices \cite{haldrup_acsnano_2016} when their dimensions become comparable to the Debye length \cite{vanderHeyden_nl_2006,vanderHeyden_nl_2007,gillespie_nl_2012}.

This growth in the potential technological applications of DLs has also spurred a growth in theories and computational techniques to describe them \cite{bikerman_1942,blum_jsp_1976,blum_jpc_1977,henderson-cpl-63-381-1979,Outhwaite19801388,Bhuiyan19811271,lozadacassou-jcp-77-5150-1982,lozadacassou-jpc-87-2821-1983,plischke-jcp-88-2712-1988,rosenfeld-prl-63-980-1989,kierlik-pra-42-3382-1990,yteran_jcp_1990,kierlik-pra-44-5025-1991,rosenfeld-jcp-98-8126-1993,rosenfeld-pre-55-4245-1997,groh-pre-57-6944-1998,boda-jcp-116-7170-2002,gillespie_jpcm_2002,gillespie_pre_2003,dicaprio-ea-2003,dicaprio-mp-2003,pizio-jcp-121-11957-2004,gillespie-jpm-17-6609-2005,reszko-zygmunt-jcp-122-084504-2005,valisko-jpcc-111-15575-2007,goel_mp_2009,henderson_jpcb_2011,Bazant_2011,wang_jpcm_2011,pizio_jcp_2012,frischknecht_jcp_2014,jiang_jcp_2014,jiang_pccp_2014,zhou_jcp_2014,kim_mp_2015,ma_jcp_2015,yang_jcp_2015,roth_jpcm_2016}.  
The need for new theories comes from the fact that the GCS theory treats ions as dilute point charges that interact through a mean electrostatic potential, an assumption that breaks down when ion size and non-mean field electrostatics produce correlations between ions \cite{torrie-jpc-1982-4,boda-jcp-116-7170-2002,henderson-pccp-11-3822-2009}.  
These correlations can manifest themselves as oscillations in the density profile of ions near electrode and in charge inversion where a layer of co-ions forms behind the first layer of counterions and can change the sign of the electrostatic potential near the electrode.
Such phenomena have been unraveled through computer simulations \cite{torrie-cpl-65-343-1979,torrie-jcp-1980-1,valleau-jcp-1982-3,torrie-jpc-1982-4,valleau-jcp-1984-5,vanmegen-jcp-1980,snook-jcp-1981,lozadacassou-jcp-77-5150-1982,lozadacassou-jpc-87-2821-1983,boda-jcp-116-7170-2002,gillespie_jpcm_2002,valisko-jpcc-111-15575-2007,zhou_jcp_2014}.
As our new technologies progress, they are encountering these correlations and have started to exploit them \cite{loessbergzahl_ac_2016}.

One problem, however, has been that there is no systematic data available against which to test new theories, something that is important as some well-established theories of DLs are qualitatively incorrect \cite{gillespie_mfnf_2014}.  
Here, our goal is to address this gap by providing high quality Monte Carlo (MC) simulation data for the primitive model of ions over a wide range of ion sizes, valences, and concentrations.  
The ion and electrostatic potential profiles we provide for download in \href{https://figshare.com/s/7b92500ab07e02c14372}{Supporting Information} can be used directly to better understand ion correlations and capacitance in general as well as in special laboratory and industrial procedures like electrophoresis \cite{khair_jfm_2009}.  
This data can also be used to check the range of validity of new theories of DLs so that creators and users of these theories can have confidence in when they are and are not correct.

Moreover, this data explicitly provides the capacitance of the DL in the primitive model for the cases we study since we provide the entire (discretized) electrostatic potential profile; the voltage holding the charge per unit area in the DL (whose total is negative of the surface charge because all simulations have well-established baths in the middle of the simulation cell) is the potential at the wall.

The primitive model is the simplest model of ions to include size by modeling them as charged hard spheres in a background dielectric that approximates the solvent.  
While all-atom simulations with explicit waters using molecular dynamics would be the ideal (see the review of Spohr \cite{spohr-ea-49-23-2003} and references therein), computing power and quality of the force fields have not reached the point where one can do a systematic study like the one here.  
That being said, the primitive model has been used for more than 40 years \cite{waisman-lebowitz}, and it has provided significant insight into DLs and reproduced many experimental results, including those with ion correlations like charge inversion \cite{he-jacs-131-5194-2009,gillespie_jcis_2011,hoffmann_l_2013}.

Since the seminal papers of Torrie and Valleau \cite{torrie-cpl-65-343-1979,torrie-jcp-1980-1,valleau-jcp-1982-3,torrie-jpc-1982-4,valleau-jcp-1984-5} who developed the Grand Canonical Monte Carlo (GCMC) technique for the DL geometry, MC simulations of the primitive model DL have been the gold standard because of the relatively few assumptions and approximations they use.  
While the ion density profiles have statistical fluctuations due to finite sampling, we have tried to minimize this by running long simulations.  
Our intent is to provide a comprehensive, high-quality database for as many cases as can realistically be dealt with GCMC or canonical simulations.

\section*{Model and method}
\label{sec:model}
\addcontentsline{toc}{section}{Model and method} 

The electrolyte is modeled as a mixture of charged hard spheres immersed in a solvent represented as a dielectric continuum of dielectric constant $\epsilon=78.5$.
The pair potential describing the interaction between two ions is then 
\begin{equation}
u_{ij}(r) =
\left\lbrace 
\begin{array}{ll}
\infty & \quad \mathrm{for} \quad r<R_{i}+R_{j} \\
 \dfrac{e^{2}}{4\pi\epsilon_{0}\epsilon} \dfrac{z_{i}z_{j}}{r} & \quad \mathrm{for} \quad r \geq R_{i}+R_{j}\\
\end{array}
\right. 
\label{eq:uij}
\end{equation} 
where $z_{i}$ and $R_{i}$ are the valence and radius of ionic species $i$, $e$ is the elementary charge, $\epsilon_{0}$ is the permittivity of vacuum, and $r$ is the distance between two ions.
Ion diameters are denoted by $d_{i}=2R_{i}$.

We used the GCMC method \cite{torrie-jcp-1980-1} to simulate this primitive electrolyte model (a misleading term; this model is simple, but far from being ``primitive'') confined between two charged hard walls at $-H$ and $H$ along the $z$-dimension.
The interaction potential between such a charged hard wall and an ion is
\begin{equation}
v_{ik}(|z|)=
\left\lbrace 
\begin{array}{ll}
\infty & \quad \mathrm{for} \quad |z|<R_{i} \\
 -\dfrac{z_{i}e\sigma_{k}|z|}{2\epsilon_{0}\epsilon} & \quad \mathrm{for} \quad |z| \geq R_{i}
\end{array}
\right. 
\label{eq:vi}
\end{equation}
where $|z|$ is the distance of the ion from the $k$th surface.
The surface charges, $\sigma_{k}$, are equal in magnitude and opposite in sign on the left and right walls.
In our simulations, the left wall is always the negative one.

The cross section of our finite simulation cell in the $x,y$ dimensions is a $L\times L$ square.
Periodic boundary conditions are applied in these dimensions.
The effect of the periodic images of the ions in the central cell are taken into account with a modified version \cite{boda-jcp-109-7362-1998} of the charged sheet method of Torrie and Valleau \cite{torrie-jcp-1980-1}.

The GCMC method includes random insertion/deletion of neutral ion clusters ($\nu_{+}$ cation and $\nu_{-}$ anions, where $\nu_{+}$ and $\nu_{-}$ are the stoichiometric coefficients).
For this, we need to know the mean chemical potential of the salt, $\mu_{\pm}=(\nu_{+}\mu_{+}+\nu_{-}\mu_{-})/(\nu_{+}+\nu_{-})$.
This was determined with the Adaptive GCMC method of Malasics et al. \cite{malasics-jcp-128-124102-2008,malasics-jcp-132-244103-2010} for a prescribed salt concentration denoted by $c$.
In our study, we deal with 1:1, 2:1, and 3:1 electrolytes, so the salt concentration is equal to the cation concentration.
The temperature was 298.15 K throughout this study.

The raw data obtained from the simulations are the density profiles, $c_{i}(z)$, computed from the average number of ions in layers of width $\mathrm{d}z$ divided by the volume of the layer ($L^{2}\mathrm{d}z$).
For convenience, we report our results for the density profiles in unit of M (mol/dm$^{3}$), so we call them concentration profiles.

The electric field and the potential were computed from solving Poisson's equation \cite{boda-hjic-41-125-2013}.
The electric field was obtained by integrating the charge profile, $q(z)=e\sum_{i}z_{i}c_{i}(z)$.
The integration constant was obtained from assuming that the electric field is zero outside the solution domain.
The electrostatic potential was obtained by integrating once more.
The integration constant was chosen so that $\Phi(-H)=0$.
Then, we spatially averaged the potential over the central part of the simulation cell (around $z=0$) and shifted the $\Phi(z)$ profile with this value.
In the simulations, we took care that a bulk region formed in the middle of the simulation cell, where $c_{i}(z)$ and $\Phi(z)$ are close to constant (this is not always easy, at low concentrations, for example). 
The result is the potential with respect to its value in the bulk.

\begin{figure}[t]
\begin{center}
\scalebox{0.53}{\includegraphics*{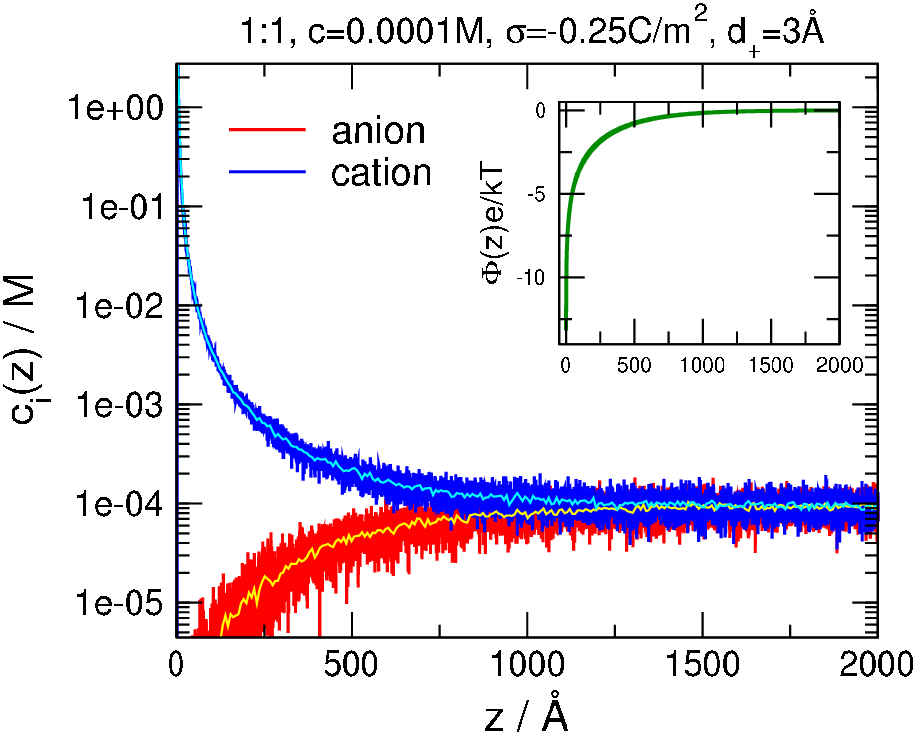}}
 \end{center}
\caption{Concentration profiles for a low concentration and high surface charge to demonstrate large noise and division between a high-density (near the electrode) and a low-density (far from the electrode) region. According to this sharp division, the electrostatic potential drops abruptly near the electrode (see the inset). Smoothed profiles are also shown with yellow and cyan.
}
\label{Fig1}
\end{figure}

The width of the bin was $\mathrm{d}z=0.25$ {\AA} in every simulation. 
This value fit an integer number times into the ionic radii used in this work, which makes the integration of the charge profiles more straightforward.
This small value, however, results in extremely noisy $c_{i}(z)$ profiles for low concentrations (see Fig \ref{Fig1}).
These profiles could be smoothed (by averaging into larger bins, for example, as shown in Fig \ref{Fig1}), but we report the raw data because they contain all the information provided by the simulations.
We trust the user to smooth the profiles.
Despite this noise due to the fine grid, the electric field and potential profiles are smooth (see inset of Fig \ref{Fig1}).

The simulations are long enough (200 millions sampled configurations, namely, attempted insertion/deletions or ion displacements) so potential profiles do not change considerably after running longer simulations.

When GCMC neutral particle group insertion/deletions fail to establish a bulk concentration (their acceptance ratio can be far below 0.1\% in the most difficult cases), we used canonical simulations. 
Their advantage is that we just displace the ions individually within the simulation box, and the acceptance ratio of that kind of MC step is much higher. 
The disadvantage is that we need to iteratively recalculate the dimensions of the simulation cell to achieve the desired bulk concentration. 
The GCMC process is more convenient (if works), while the canonical iterative process is cumbersome, but sometimes it is the only way (e.g., 3:1 electrolyte with $c=1$ M).

\begin{figure}[t]
 \begin{center}
\scalebox{0.52}{\includegraphics*{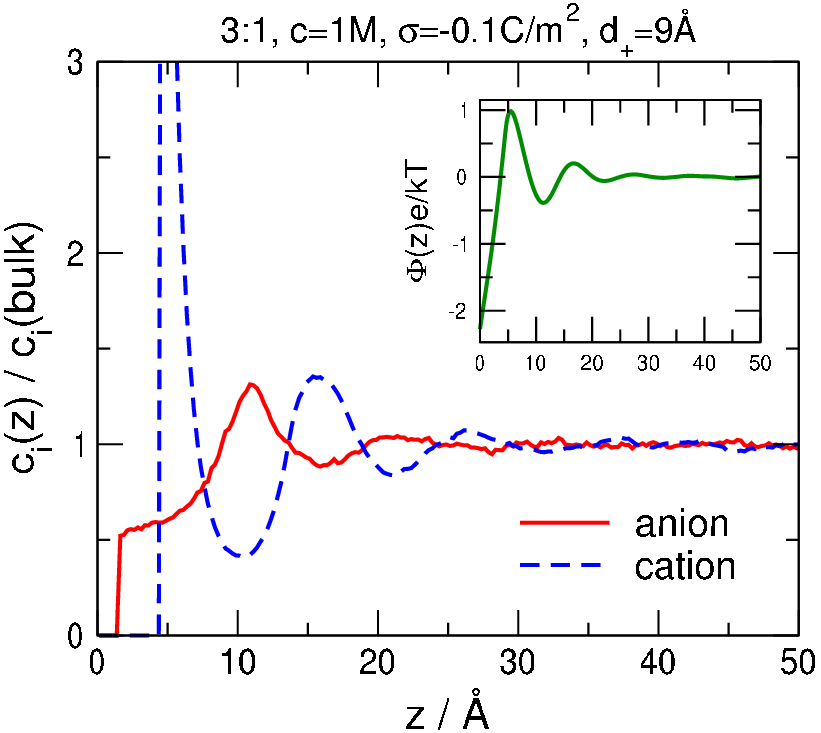}}
 \end{center}
\caption{Concentration profiles for a high concentration and 3:1 electrolyte to demonstrate the noise associated with problematic sampling (low acceptance ratio). Concentration profiles are normalized to 1 with the bulk value. Strong layering due to asymmetries both in ion charge and size can be observed that results in charge inversion as also seen from the electrostatic potential profile (see the inset).
}
\label{Fig2}
\end{figure}


\section*{Results}
\label{sec:results}
\addcontentsline{toc}{section}{Results} 

We performed simulations for an extended range of state variables.
The valence and diameter of the anion were fixed at the values $z_{-}=-1$  and $d_{-}=3$ {\AA}, respectively.
We changed the cation attributes by varying its valence ($z_{+}=1$, 2, and 3) and diameter ($d_{+}=1.5$, 3, 6, and 9 {\AA}).

We varied the bulk concentration in steps by orders of magnitude ($c=0.0001$, 0.001, 0.01, 0.1, and 1 M) so we cover a wide range, from extremely dilute to quite concentrated.
The surface charge is varied in finer steps up to $|\sigma|=0.1$ C/m$^{2}$ (0, 0.02, 0.04, 0.06, 0.08 and 0.1 C/m$^{2}$). 
This makes it possible to compute capacitance at zero electrode charge (data not provided).
We also tested quite high surface charges up to $|\sigma| =0.5$ C/m$^{2}$ (0.175, 0.25, 0.375, and 0.5 C/m$^{2}$), where it was possible.

Simulations were straightforward mostly, but in several cases it was difficult to establish a bulk region in the middle of the simulation cell.
Therefore, we do not report results for the following cases:
\begin{itemize}
 \item $c=0.0001$ M and $c=0.001$ M for the 3:1 case
 \item $c=0.0001$ M for the 2:1 case
 \item $c=0.0001$ M for the 1:1 case at high surface charges
\end{itemize}
\begin{figure*}[t]
 \begin{center}
\rotatebox{0}{\scalebox{0.59}{\includegraphics*{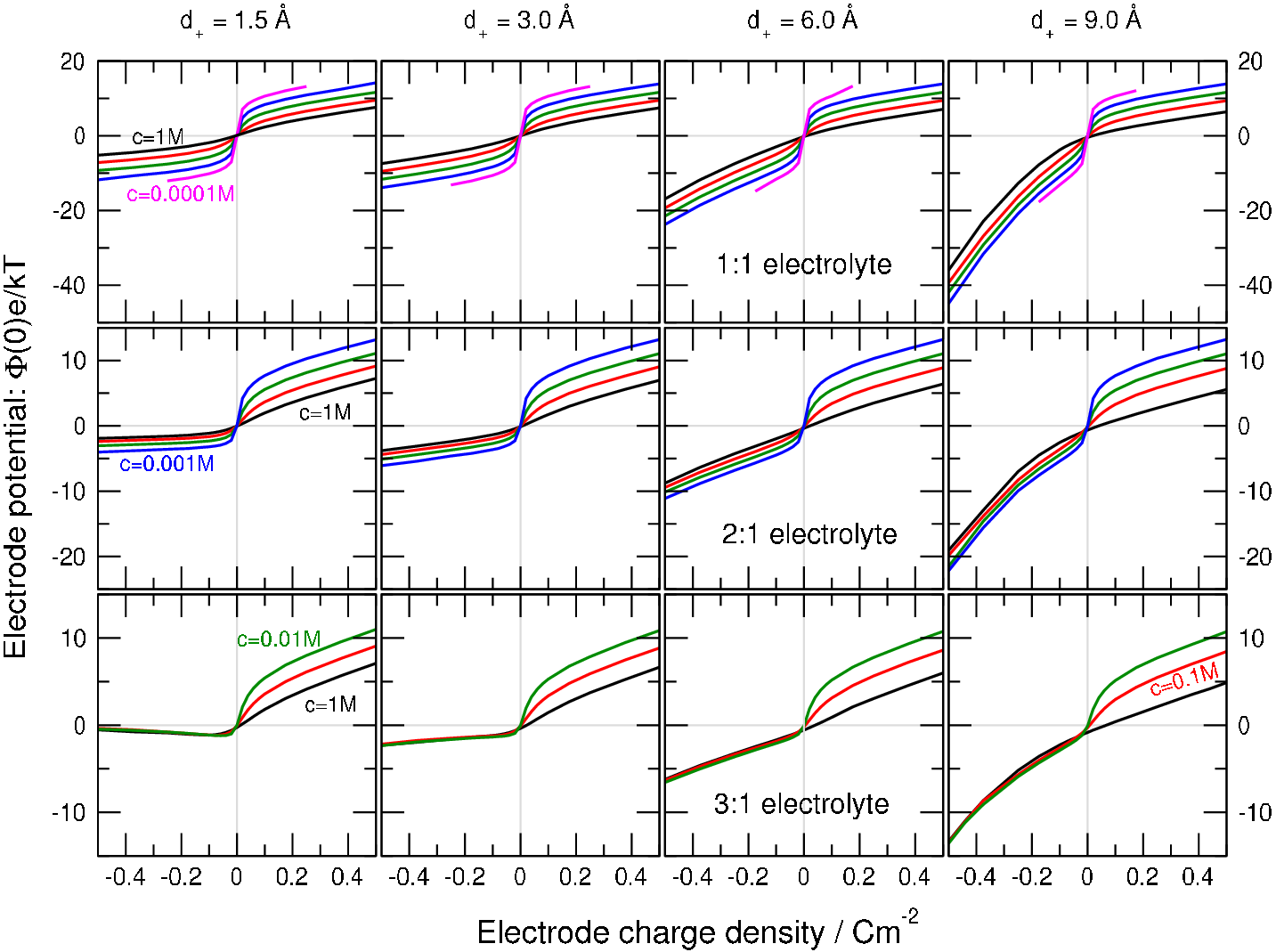}}}
 \end{center}
\caption{Electrostatic potentials at the electrode as functions of the electrode charge.
The three rows of the figure refer to various cation valences (1:1, 2:1, and 3:1 electrolytes), while the four columns refer to various cation diameters ($d_{+}=1.5,$ 3, 6, and 9 {\AA}). 
The lines in a given panel refer to different concentrations: $c=0.0001$ (magenta), 0.001 (blue), 0.01 (green), 0.1 (red), and 1 M (black).
}
\label{Fig3}
\end{figure*} 
These cases were difficult for the following reasons:
\begin{itemize}
 \item Low concentrations with their large Debye lengths are complicated because the simulation cell must be long enough ($H\approx2000$ {\AA}) in order to fit the wide DL into it. 
Large cross sections, $L$, must also be used in order to include enough ions into the cross section so we can sample ion correlations not only in the $z$, but also in the $x,y$ dimensions. 
Sampling of possible ion configurations in this huge space is difficult when we have an otherwise dilute system.
 \item It is especially difficult in the case of 2:1 and 3:1 electrolytes, where we insert/delete 3 or 4 ions instead of just 2. The acceptance ratio of the insertions/deletions, therefore, is very small especially at high concentrations. 
In most cases, however, we were able to collect enough sample from insertions/deletions except for 3:1 electrolytes at $c=1$ M using large cations ($d_{+}=6$ and 9 {\AA}). In these cases, we used canonical simulations.
 \item The other especially difficult case is when the surface charge is large because the system is divided into a very high density (for the ions) region near the electrode and a low density region in the bulk. 
\end{itemize}
Fig \ref{Fig1} illustrates the problematic case associated with low $c$ and high $\sigma$.

The other problematic case is large concentration ($c=1$ M) with multivalent cations (especially $z_{+}=3$).
In this case, the fine noise associated with low density is replaced by a coarser noise associated with high density and low acceptance ratios of particle moves (Fig \ref{Fig2}).
In these cases, a few percent of error in the reproduced bulk densities can occur. 
These errors have little effect on the potential profiles.

Also, a long range ordering can be observed (oscillations in profiles that extend far from the wall) that is sometimes difficult to distinguish from noise (Fig \ref{Fig2}).
The structure of the DL near the electrode is, however, well established; the potential profiles, therefore, can be calculated accurately.
We can state that, in general, the potential is dominated by the region near the electrode, while the tail of the diffuse layer reaching into the bulk has a smaller effect.

\begin{figure*}[t]
 \begin{center}
\rotatebox{0}{\scalebox{0.59}{\includegraphics*{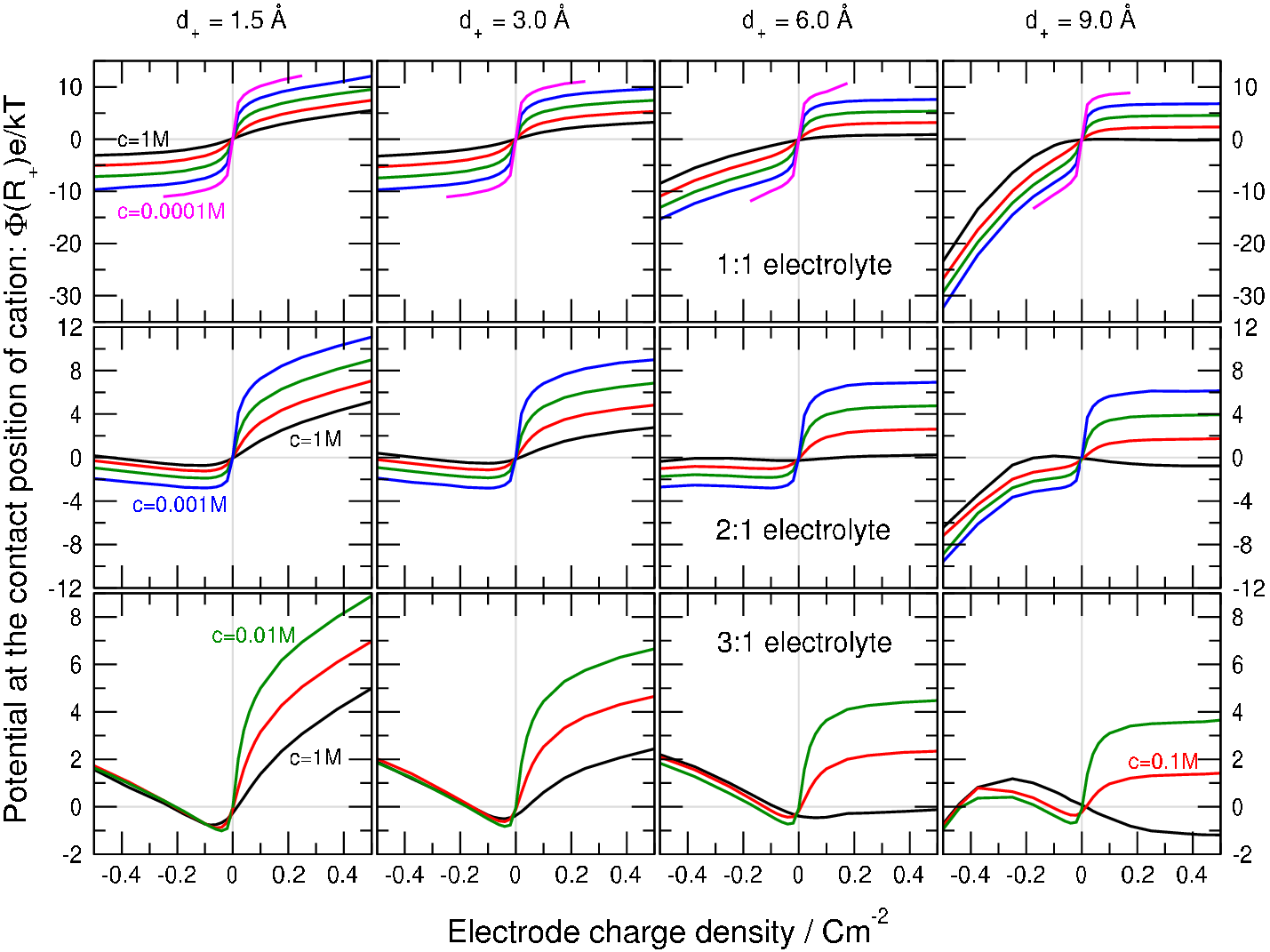}}}
 \end{center}
\caption{Electrostatic potentials at the contact position of the cation ($z=d_{+}/2$) as functions of the electrode charge.
The three rows of the figure refer to various cation valences (1:1, 2:1, and 3:1 electrolytes), while the four columns refer to various cation diameters ($d_{+}=1.5,$ 3, 6, and 9 {\AA}). 
The lines in a given panel refer to different concentrations: $c=0.0001$ (magenta), 0.001 (blue), 0.01 (green), 0.1 (red), and 1 M (black).
}
\label{Fig4}
\end{figure*} 

The potential at the electrode is shown in Fig \ref{Fig3} as it is a quantity of basic interest.
The potential at the cation radius (Fig \ref{Fig4}) is closely related to the otherwise poorly-defined zeta potential that is supposed to be the potential at the slip plane that separates immobile ions attached to the electrode from those that are mobile in the diffuse layer. 
This value, therefore, has practical importance.
A few conclusions on the basis of these figures follows:
\begin{itemize}
 \item Very significant (down to $-40kT/e$) electrode potentials can be reached using large monovalent cations. 
This has relevance when we want to design nanofluidic devices turning hydrostatic energy into electrical power \cite{gillespie_nl_2012} or when sensing surface charge changes \cite{friedrich_sab_2016}.
 \item For the 3:1 systems, the electrode potential is independent of concentrations for negative surface charges because the trivalent cations adsorbed closely to the electrode dominate the DL. The diffuse layer has very small effect.
 \item In the 3:1 case, we see negative differential capacitance for $d_{+}=1.5$ {\AA} (Fig \ref{Fig3}).
 \item An interesting maximum appears in Fig \ref{Fig4} for $d_{+}=9$ {\AA}, which is the result of the balance between opposing effects of large ionic charge (increasing field at the ion's surface) and large ionic size (decreasing field at the ion's surface). Fig \ref{Fig2} shows the effect of these competing asymmetries in the concentration and potential profiles. 
\end{itemize}


\section*{Conclusions }
\label{sec:conc}
\addcontentsline{toc}{section}{Conclusions} 

We have produced systematic simulation results for the classic model of the DL over a wide range of parameters.
Some of these parameters (small concentration, large electrode charge, large ionic charge, large ionic size) have historically posed a challenge for reasons we discussed.
Now, however, there is sufficient computing power and patience to overcome many of these challenges.
There is also a compelling scientific need.
By making this data freely available, research groups can use it directly in their work or to test new theories. 

\section*{Supporting information}
\addcontentsline{toc}{section}{Supporting information} 

\paragraph*{S1 File.}
\label{S1_File}
{\bf Supporting pdf.} Tabulates basic parameters and the potential values shown in Figs \ref{Fig3} and \ref{Fig4}. Explains nomenclature for directory and file names in the compressed subdirectory system in \nameref{S3_File}. Available from 
\href{https://figshare.com/articles/Document_containing_model_description_parameters_results_and_description_of_the_compressed_file_containing_profiles/5436238}{\ttfamily \underline{figshare.com}}

\paragraph*{S2 File.}
\label{S2_File}
{\bf Supporting sheets.} Tabulates the same basic parameters and the potential values shown in \nameref{S1_File} in sheets. Available from 
\href{https://figshare.com/articles/Sheets_containing_simulation_parameters_and_results/5436241}{\ttfamily \underline{figshare.com}}

\paragraph*{S3 File.}
\label{S3_File}
{\bf Supporting database.} A compressed file that is a zipped subdirectory system organized into a tree structure. 
This subdirectory system contains all the concentration, electric field, and electrostatic potential profiles along with basic simulation parameters for each run. Nomenclature for directory and file names is given in the \nameref{S1_File}. Available from 
\href{https://figshare.com/articles/Compressed_subdirectory_system_containing_all_simulation_data_including_profiles/5436244}{\ttfamily \underline{figshare.com}}

\section*{Acknowledgments}

DB and MV gratefully acknowledge the financial support of the National Research, Development and Innovation Office -- NKFIH K124353 in the framework of ERA Chemistry. 
This material is based upon work supported by the National Science Foundation under Grant No. 1402897 (to D.G.).
Present article was published in the frame of the project GINOP-2.3.2-15-2016-00053 (``Development of engine fuels with high hydrogen content in their molecular structures (contribution to sustainable mobility)'').
Supported by the UNKP-17-4 (to MV) New National Excellence Program of the Ministry of Human Capacities.
We appreciate the help of Adelina Voukadinova in thoroughly checking the database.
We thank Zolt\'an Hat\'o for his help with building the database.

\phantomsection



\end{document}